# First-Order Structural Change Accompanied by Yb Valence Transition in YbInCu$_4$

Satoshi Tsutsui [1,*], Kunihisa Sugimoto [1], Ryoma Tsunoda [2], Yusuke Hirose [3], Takeshi Mito [4], Rikio Settai [3] and Masaichiro Mizumaki [1,**]

[1] *Japan Synchrotron Radiation Research Institute, SPring-8, Sayo, Hyogo 679-5198, Japan*
[2] *Graduate School of Science and Technology, Niigata University, Niigata 950-2181, Japan*
[3] *Department of Physics, Niigata University, Niigata 950-2181, Japan*
[4] *Graduate School of Material Science, University of Hyogo, Kamigori, Hyogo 678-1297, Japan*

Corresponding authors: * satoshi@spring8.or.jp, ** mizumaki@spring8.or.jp

PACS No.:

**Abstract**

A diffraction experiment using a high energy x-ray was carried out on YbInCu$_4$. Below the Yb valence transition temperature, the splitting of Bragg peaks was detected in higher-order reflections. No superlattice reflections accompanying the valence ordering were found below the transition temperature. These experimental findings indicate that a structural change from a cubic structure to a tetragonal structure without valence ordering occurs at the transition temperature. Such a structural change free from any valence ordering is difficult to understand only in terms of Yb valence degrees of freedom. This means that the structural change may be related to electronic symmetries such as quadrupolar degrees of freedom as well as the change in Yb valence.

Valence degrees of freedom are always coupled with structural degrees of freedom because the ionic radius is changed by a change in the valence state. When valence ordering occurs in materials, the space group of their crystal structure can change because of the alignment of the ions with valence states. Meanwhile, when a valence change occurs without valence ordering, the unit cell volume is expected to change owing to the change in the ionic radius but the symmetry of the crystal structure is not expected to be changed by a valence transition. If a valence transition with a symmetric change but no valence ordering occurs, the discovery of such a valence transition would shed new light on the valence transition compounds.

Valence states of rare-earth atoms are often strongly related to the unit cell volume in rare-earth intermetallics. In such a case, the valence states of rare-earth atoms in rare-earth intermetallic compounds can be estimated from the lattice constant and/or thermal expansion on the basis of lanthanide contraction. A typical example is the temperature dependence of valence state and thermal expansion in $SmB_6$ [1, 2, 3]. Its unit cell volume is related to the Sm valence without any symmetric changes. This means that crystallographic investigation, particularly determination of the lattice constant or unit cell volume of a material, is helpful for understanding the valence state of rare-earth atoms in rare-earth intermetallic compounds. A similar interpretation can be applied to the valence-ordering system of $Yb_4As_3$ [4]. The valence ordering leads to a change in the crystallographic symmetry owing to the difference between the ionic radii of $Yb^{2+}$ and $Yb^{3+}$. From these viewpoints, crystallographic investigation is a useful approach for discussing valence transition and/or ordering.

The series of Yb$T$Cu$_4$ (T = Ag, Au, Cd, Mg, Tl, and Zn) has been recognized as valence-fluctuating compounds [5]. Only $YbInCu_4$ exhibits a valence transition in the series of Yb$T$Cu$_4$ compounds [6, 7]. The valence transition in $YbInCu_4$ has been examined by many techniques. The temperature dependence of the Yb valence state was investigated by the synchrotron radiation (SR) techniques of X-ray absorption spectroscopy, X-ray emission spectroscopy, and photoemission spectroscopy measurements [8, 9, 10, 11]. Thermal expansion accompanied by the change in the Yb valence was discussed using neutron diffraction experiments [12, 13, 14]. Both sets of experimental results suggest the absence of any structural transitions accompanied by the Yb valence transition. Although a structural change was not suggested, line broadening of the Bragg reflection below the valence transition temperature was reported in Ref. [13]. Therefore, it has been believed that $YbInCu_4$ undergoes a valence transition with thermal expansion or contraction due to a change in the Yb ionic radius. Meanwhile, although the shift of the Yb valence at the transition temperature is much smaller than difference of valence between the pure $Yb^{2+}$ and $Yb^{3+}$ states [10, 11, 14], low-energy spectra obtained by inelastic neutron scattering were dramatically changed as if the Yb valence state in $YbInCu_4$ had changed from a magnetic trivalent state to a nonmagnetic divalent state [15]. Such a dramatic change in the low-energy spectra is not easy to explain by the observed small shift of the Yb valence state.

In the present work, we carried out high-energy x-ray diffraction experiment of $YbInCu_4$

using SR to clarify the Yb valence transition mechanism. A high-energy x-ray expands the Ewald sphere in diffraction experiments, which enables us to perform a more precise structural investigation using SR than by laboratory diffraction apparatus. Splitting of Bragg reflections was observed in higher-order reflections at 30 K in the present work, suggesting a change in the crystallographic symmetry, but no superlattice reflections were observed. The former indicates a change in the crystallographic symmetries accompanied by a valence transition. The latter indicates the absence of any Yb valence ordering. These results imply that ferro-type ordering due to nonmagnetic degrees of freedom occurs at the valence transition in $YbInCu_4$.

A single-crystal sample was prepared by the In-Cu flux method [16]. To determine the quality of the sample, we performed magnetic susceptibility and x-ray absorption spectroscopy (XAS) measurements. The temperature dependence of the magnetic susceptibility at an applied magnetic field of 0.5 T is shown in Fig. 1. The obtained magnetic susceptibility agrees with that in previous works on $YbInCu_4$ [16, 17, 18, 19, 20]. A sudden drop of magnetic susceptibility was found at about 40 K, suggesting a valence transition from a magnetic state to a nonmagnetic state. Meanwhile, the XAS spectra measured at BL39XU of SPring-8 are shown in Fig. 2. These spectra obtained at 3 and 280 K also agree with those in a previous work [8]. The spectra of $YbInCu_4$ shown in Fig. 2 demonstrate a Yb valence transition from an intermediate Yb valence state to a nearly pure $Yb^{3+}$ state at about 40 K.

A single crystal x-ray diffraction experiment on $YbInCu_4$ was carried out at BL02B1 of SPring-8. The SR x-ray energy delivered from a bending magnet was tuned to an energy of 27.00 keV ( = 0. 459 Å) using a Si(3 1 1) double-crystal monochromator. The beam was focused to a size of about 0.1 × 0.1 mm$^2$ (FWHM) by a sagittal focusing monochromator and a bent mirror made of Si crystal coated with Pt. A cylindrical imaging plate with a camera length of 191.3 mm was adopted for precise structure analysis of the dependence of the temperature at 30 and 300 K using a cryogenic He flowing system (XR-HR10K-S, Japan Thermal Engineering Co. Ltd.). Six frames of diffraction images in the momentum transfer range of up to about 26 Å$^{-1}$ were taken with an exposure time of 18 minutes at each temperature. The oscillation angle of the crystal in each frame was 30 degrees. All structures were solved by a direct method (*SHELXS*-97) and refined by the full-matrix least-squares technique on $F^2$ (*SHELXL*-97). We used the anomalous dispersion coefficients for structure refinement when considering the dependence of scattering cross section on x-ray energy using the original FPRIME code of Cromer [21]. These coefficients are $f'$ = -0.42010 and $f'$= 2.65478 for ytterbium, $f'$ = -2.75663 and $f''$= 0.59620 for indium, and $f'$= 0.27873 and $f''$= 0.57381 for copper.

Figure 3 shows oscillation photographs of YbInCu$_4$ taken at 30 and 280 K. Since YbInCu$_4$ exhibits a valence transition at about 40 K, any differences between the photographs taken at 30 and 300 K indicate a structural change due to the valence transition. Two evident findings were demonstrated by the photographs: one is the absence of a long-period structure; the other is a change in the crystallographic symmetry. No superlattice reflections were found in the oscillation photograph taken at 30 K. Since YbInCu$_4$ exhibits valence transition at about 40 K, the oscillation photograph taken at 30 K demonstrates the absence of long-period structures suggesting valence ordering. Instead, splitting of the reflections with a 2$\theta$ value greater than about 90 degrees was found at 30 K, whereas no such splitting was found at 300 K. Observation of the splitting of reflection spots by the domain structure in the higher-momentum-transfer region indicates a phase transition from a cubic structure to a tetragonal structure occurs at the valence transition temperature. The splitting of Bragg reflection spots in the lower-momentum-transfer region is not clear, which suggests that the change in the c/a value due to the valence transition is small. The estimated c/a value obtained from structural analyses is 1.000005 at 30 K.

The crystallographic analyses using the oscillation photographs taken at both 30 and 300 K were successful. The crystallographic parameters at 30 and 300 K were determined as shown in Table 1. These parameters reveal that the space groups are different at 30 and 300 K. The former is F-43m having cubic symmetry, and the latter is I-4m2 having tetragonal symmetry. The unit cell in the *ab*-plane at 30 K is rotated by 45 degrees from that at 300 K as shown in Fig. 4. The number of atoms per unit cell volume at 30 K is about half of that at 300 K. This means that the atomic position parameters of all the atoms are nearly identical except for a change in the c/a value when the unit cell at 30 K is defined as the same as that at 300 K. In addition, Table 1 shows that the unit cell volume at 30 K is identical to that at 300 K. Although the unit cell volume decreases with decreasing temperature in general, the present results demonstrate that the thermal expansion due to the change in the average Yb ionic radius, indicated by the results of the XAS measurements shown in Fig. 2, is nearly compensated by the thermal contraction with decreasing temperature.

The change in symmetry revealed in the present work has not been reported previously. However, the disagreement between the present work and previous studies may have been caused by the previous analyses under the assumption of cubic symmetry [12, 13, 22]. In fact, the physical properties of the sample used in the present work are nearly identical to those of the sample used in a previous work [12]: the lattice constants at 30 and 300 K in the present work agree with those estimated from the temperature dependence of 4 2 2 reflection; the temperature of the sudden drop in the magnetic susceptibility shown in Fig. 1 agrees with that of the jump in the lattice constant. In addition, even a symmetric change may agree with the present work. One possible piece of evidence for agreement with the present work is the broadening of 3 1 1 Bragg reflection obtained in a neutron diffraction experiment [13]. Even if the momentum transfer resolution in the neutron diffraction experiment was inferior to that in the present work, there are no reasons why line broadening should occur without any structural transitions at a low temperature. In this sense, the splitting of the Bragg reflection observed in the present work may have been observed as line broadening in the neutron diffraction experiment.

Structural changes accompanying the change in the crystallographic symmetry have been reported for some Yb intermetallic compounds. One is $Yb_4As_3$ and another is YbPd. In both cases, the structural changes were caused by Yb valence ordering for different reasons. $Yb_4As_3$ exhibits a structural change from I-43d to R3c at its valence transition temperature [4]. This transition can be interpreted as valence ordering along the <1 1 1> direction from the viewpoint of the ionic radius. Lowering of the crystallographic symmetry in $Yb_4As_3$ is produced by difference between ionic radii of the Yb divalent and trivalent states. Meanwhile, YbPd exhibits successive transitions from a cubic structure to a tetragonal structure [23, 24]. The ordering vector of the valence transition is $q = (1/2\ 0\ 0)$. However, valence ordering with $q = (1/2\ 0\ 0)$ is difficult to explain for a tetragonal structure in the low temperature phase. A crystallographic analysis suggests that a position shift of Pd atoms is important for interpreting the x-ray diffraction pattern in the low temperature phase [23]. This suggests that the change in the symmetry around Yb atoms is important, at least in the case of YbPd.

Similarly, the change in the symmetry around Yb atoms is crucial when considering the structural change in $YbInCu_4$. In the case of $YbInCu_4$, however, no valence ordering but a small amount of lattice distortion was observed even in the transition from a cubic structure to a tetragonal structure. When the energy gain in the electronic system due to the Jahn-Teller effect is considered, quadrupole ordering is equivalent to a structural change accompanied by lattice distortion from the viewpoint of removing electronic degeneracy. Hereafter, a phase transition with a large lattice distortion is defined as the Jahn-Teller effect and that with a small amount of lattice distortion is defined as quadrupole ordering, because distinguishing between the Jahn-Teller effect and quadrupole ordering is difficult. According to this definition, the present result implies that YbInCu$_4$ undergoes quadrupole ordering accompanied by a valence transition. While multipolar degrees of freedom in Yb$^{3+}$ ions, such as 4f quadrupolar degrees of freedom, as well as valence degrees of freedom exist in Yb compounds, the energy gain can be produced in the electronic system by removing the degeneration of electronic levels owing to the 4f quadrupolar degrees of freedom.

The level scheme of the crystal electric field in YbInCu$_4$ has been investigated previously but it was difficult to conclud without any uncertainties [15, 16, 17, 25, 26, 27]. A plausible interpretation based on previous reports suggests that the ground state is a $\Gamma_8$ quartet, the first excited state is either a $\Gamma_6$ or $\Gamma_7$ doublets, and the second excited state is the other doublet. Difference of energy between the first and second excited states is a few Kelvin, while that between the ground and first excited states is larger than 35 K. It has not yet been precisely determined which of the $\Gamma_6$ and $\Gamma_7$ state is higher. Considering the $\Gamma_8$ quartet ground state and the structural change from a cubic structure to a tetragonal structure, the expected ground state below the transition temperature is a doublet. One of the possible order parameters in YbInCu$_4$ is a quadrupolar moment. If ferroquadrupole ordering, which can exclude any long-period ordering accompanied by superlattice reflections, occurs at the valence transition temperature, electronic degeneration can be removed with little lattice distortion. In fact, in the present work a structural change from a cubic structure to a tetragonal structure with extremely small lattice distortion has been confirmed. This experimental finding infers that quadrupole ordering accompanied by a valence transition is realized in YbInCu$_4$, because the XAS results shown in Fig. 2 imply that quadrupolar degrees of freedom do not disappear above or below the Yb valence transition at about 40 K.

The experimental results shown in Figs. 2 and 3 mean that the internal energy increases owing to an increase in the unit cell volume in this system but the f-electron degrees of freedom still remain even below the transition temperature. In other words, competition is realized between structural degrees of freedom that change the unit cell volume and f-electron degrees of freedom, possibly supporting ferroquadrupole ordering in the low temperature phase of YbInCu$_4$. If the structural tetragonality in the low temperature phase is ignored because of the small deviation of the c/a value from unity, the order parameter at the valence transition temperature is $q$ = (0 0 0). In addition, the temperature factors obtained in a previous diffraction experiment suggest there are no anomalies in the phonon dispersion relations in $YbInCu_4$ [22]. However, the effects of f-electrons, particularly Yb valence state, on the dynamical structure cannot be excluded in this discussion. Also, quadrupolar susceptibilities are usually coupled with elastic constants. In other words, the f-electron effects accompanied by the Yb valence transition are not obvious. The phonon dispersion relations in $YbInCu_4$ are worth investigating.

In summary, we found a symmetric change in the crystal structure of $YbInCu_4$ between the high and low temperature phases in the present work. The crystal structure in the low temperature phase does not have any long-period ordering, suggesting absence of Yb valence ordering. The symmetric change due to the valence transition is caused by a change from a cubic structure to a tetragonal structure. This is difficult to explain in terms of a valence transition free from any valence ordering. In addition, the lattice distortion due to the structural change is small. More precise measurements from a macroscopic viewpoint are required to conclude that this crystal transition is caused by a structural transition accompanied by 4f-electron degrees of freedom. However, these experimental finding may support ferroquadrupole ordering accompanied by a small amount of lattice distortion rather than a Jahn-Teller type transition accompanied by a large amount of lattice distortion in the ground state of $YbInCu_4$. The present work implies that a novel type of ferroquadrupolar ordering may occur at the valence transition temperature in $YbInCu_4$.

**Acknowledgements**
The authors appreciate T. Hasegawa and H. Harima for their helpful discussions. The present work was carried out under the approval of JASRI (Proposal Nos. 2015A0046, 2015A1492, 2015A2061). It was financially supported by Grants-in-Aid for the Scientific Research (B) (15H03697) and (C) (26460057) from the Japan Society of Promotion of Science.

**Table**

Table 1. Crystallographic parameters of YbInCu$_4$ at 30 and 300 K.

| Temperature (K) | | 30 | 300 |
|---|---|---|---|
| Space Group | | I-4m2 (#119) | F-43m (#216) |
| Lattice constant (Å) | | $a$ = 5.0618(5), $c$ = 7.1585(4) | $a$ = 7.1586(3) |
| Yb | (x, y, z) | 0.00, 0.50, 0.25 | 0, 0, 0 |
| | Wyckoff position | 2$c$ | 4$a$ |
| In | (x, y, z) | 0.0, 0.0, 0.0 | 0.75, 0.75, 0.75 |
| | Wyckoff position | 2$a$ | 4$d$ |
| Cu | (x, y, z) | 0.25041(15), 0.0, 0.37475(9) | 0.37515(3), 0.37515(3), 0.37515(3) |
| | Wyckoff position | 8$i$ | 16$e$ |

**Figure Captions**

Fig. 1. Temperature dependence of Magnetic susceptibility of YbInCu$_4$ at a magnetic field of 0.5 T.

Fig. 2. Yb L$_{III}$-edge XAS spectra of YbInCu$_4$ obtained at 3 and 280 K. Open (closed) circles denote data obtained at 3 (280) K.

Fig. 3. Oscillation photographs of YbInCu$_4$ taken at 30 K (upper panel) and 300 K (lower panel). The vertical direction corresponds to the [1 1 1] direction, which is the rotation axis of the sample in the present experiment.

Fig. 4. Crystal structure of YbInCu$_4$ in (a) high temperature phase and (b) low temperature phase.

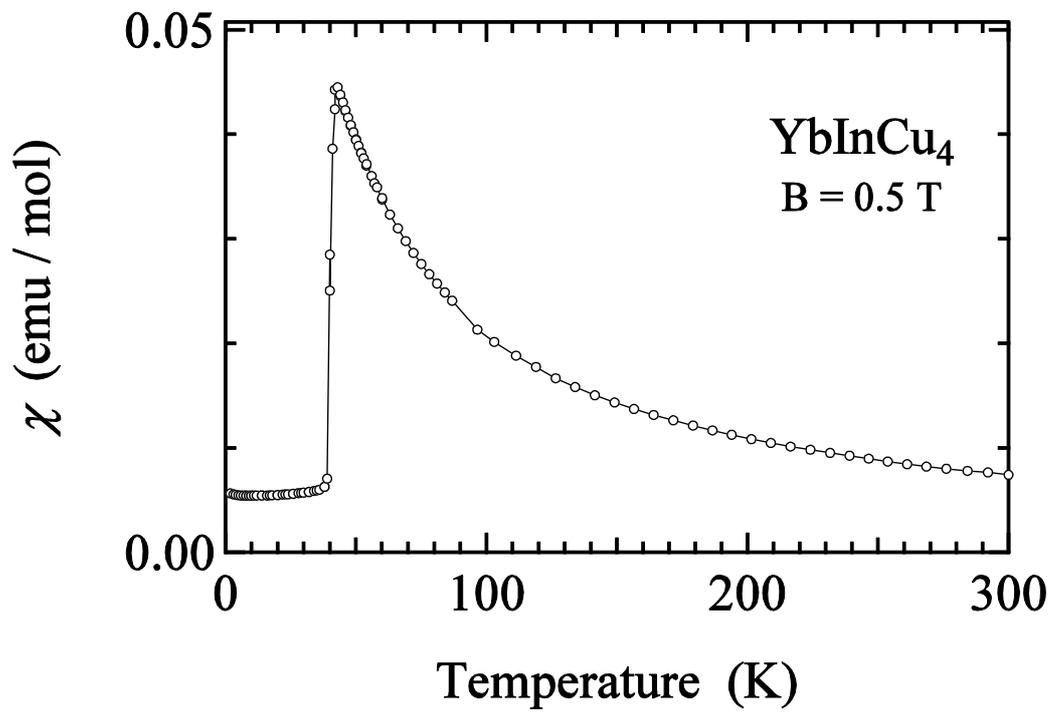

Fig. 1

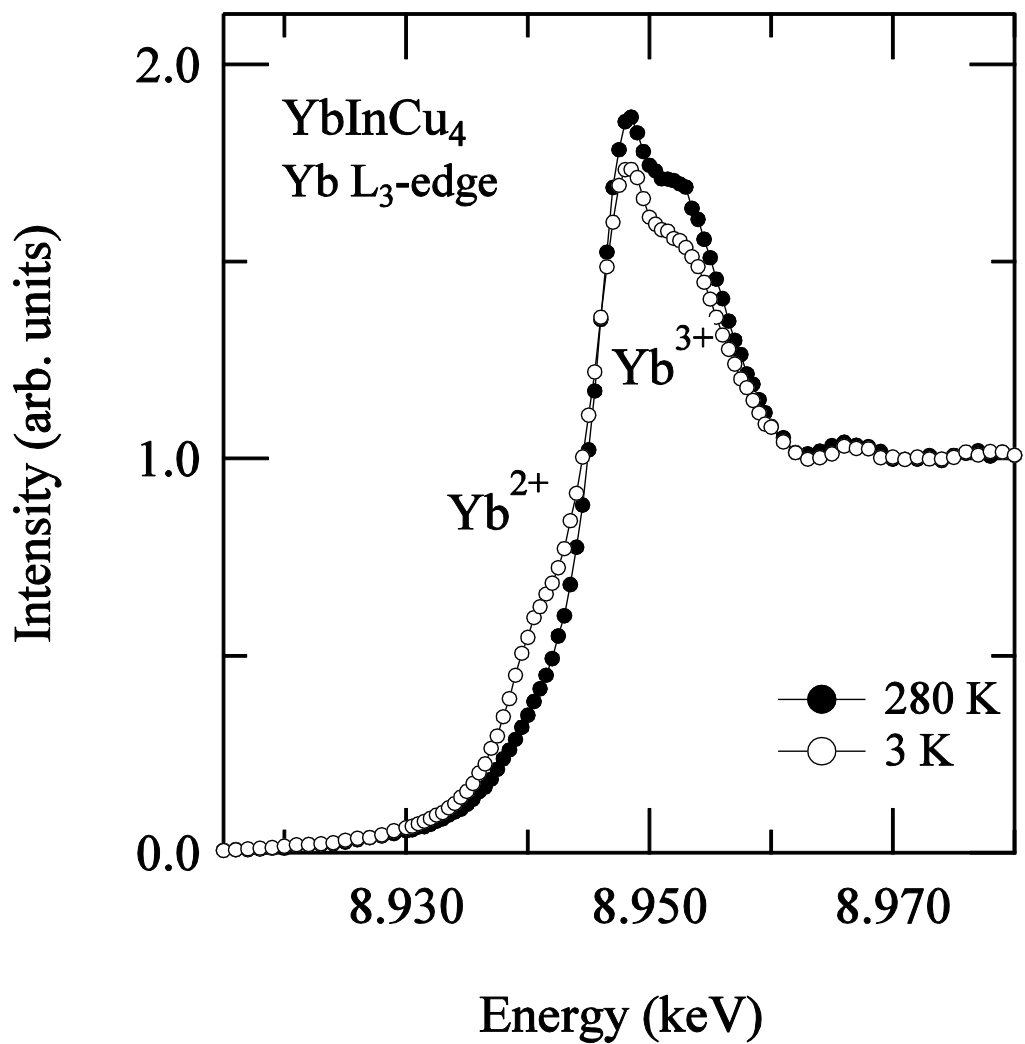

Fig. 2

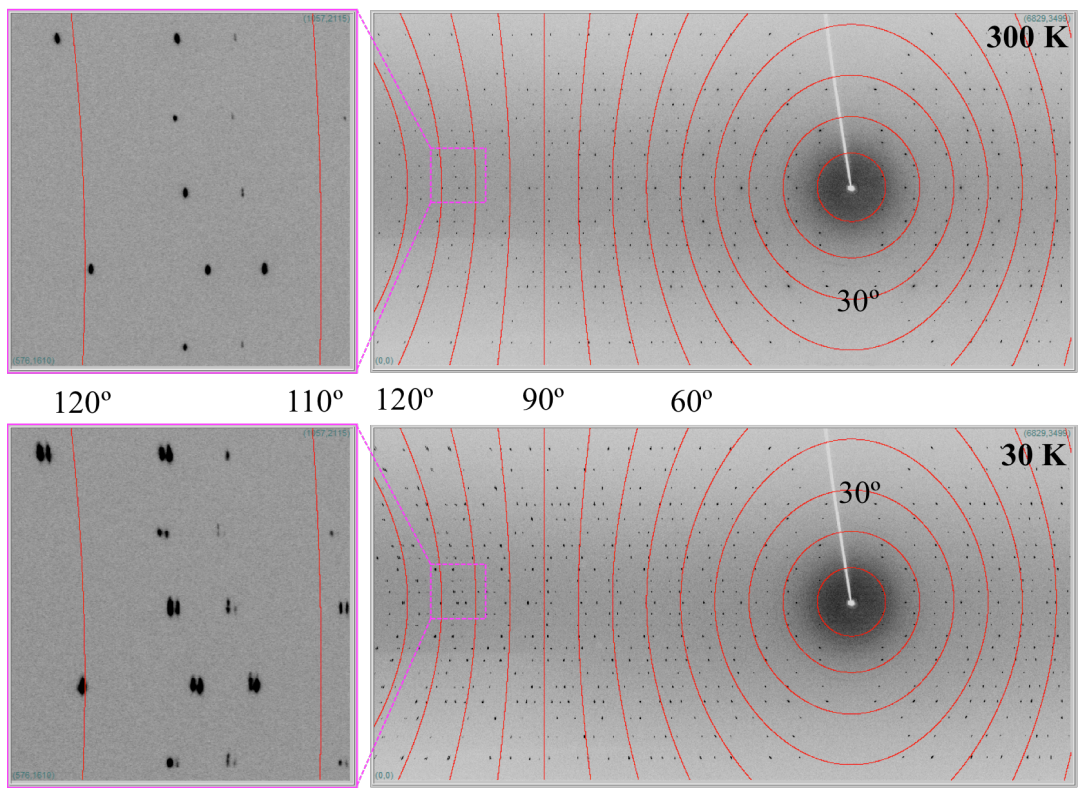

Fig. 3
13

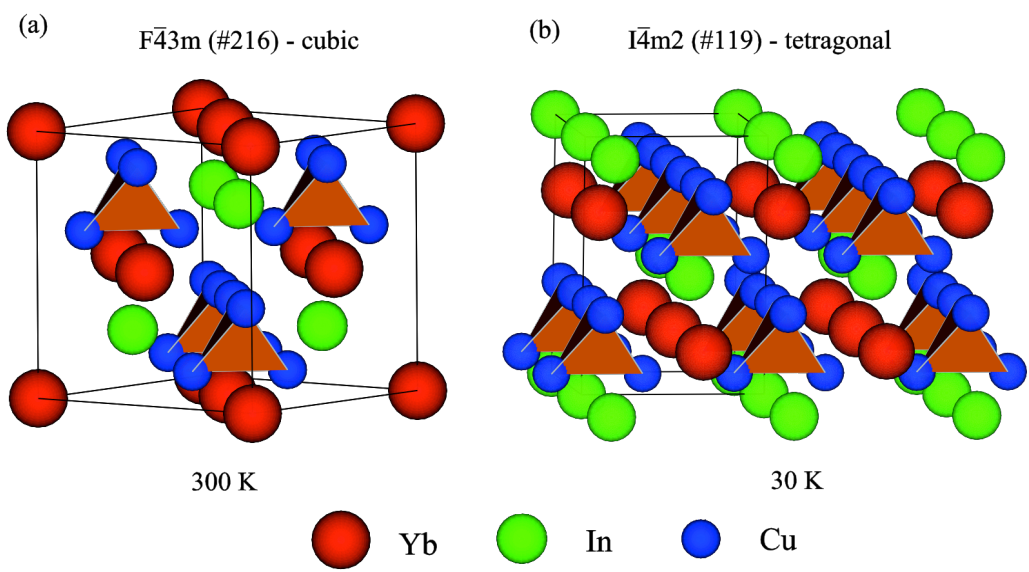

Fig. 4